# Plant Growth Estimation with a Camera-Based Vegetation Index Mapping System for Agricultural Ground Vehicles ⋆

**Lukas Pindl** * **Michael Maier** * **Timo Oksanen** *,**

* *Professorship for Agrimechatronics, Technical University of Munich, Munich, Germany (email: lukas.pindl@tum.de, michael82.maier@tum.de, timo.oksanen@tum.de)*
** *Munich Institute of Robotics and Machine Intelligence (MIRMI), Munich, Germany*

**Abstract:** This work presents a camera-based sensor for mapping plant growth over an agricultural field. The sensor can be used on ground-based vehicles like a tractor and relies on RTK-GNSS to correctly merge many multispectral images onto one map. NDVI is used as an index to estimate plant growth, but the approach can be used with other indices as well depending on the camera. The images from the camera are projected onto the estimated ground plane using perspective projection. This computationally simple approach allows for real-time processing of all images on the field even with low-end hardware. The results are compared to a commercially established vehicle-mounted sensor. Absolute values are hard to compare, but both sensors show similar trends. The camera-based approach also allows for filtering of ground and non-ground areas, potentially reducing the impact of crop density on the average measured NDVI.

*Keywords:* Computer vision in agriculture, Sensing and perception in agriculture, Vegetation Indices, Plant Monitoring, Nitrogen Use Efficiency

## 1. INTRODUCTION

### 1.1 Motivation

Reducing nitrogen usage is a core goal of agricultural politics in many countries worldwide (Shen et al. (2024), Geupel et al. (2021), EPA (2019)). Lowering the overall amount of nitrogen applied without losing yield necessitates a higher nitrogen use efficiency (**NUE**). One technique to decide where to put more or less fertilizer is using vegetation indices to identify areas of higher or lower potential yield on the field. These indices typically use optical sensors to estimate the health and growth state of plants (Rouse et al. (1974), Barnes et al. (2000)). The Normalized Difference Vegetation Index (**NDVI**), for example, is commonly computed from multispectral camera images mounted on satellites or drones. For drones, many independent images are typically taken with a rough GPS position and then later stitched together using, e.g., Open Drone Map (OpenDroneMap Authors (2020)). There also exist sensors that can be mounted to ground vehicles to accurately map plant growth in the field independent of the weather. However, these sensors typically have a low resolution, using only a single optical sensor. They are also highly expensive, making them interesting only for large-scale farms. We propose the use of camera back-projection to map an area directly from a multispectral camera mounted to a tractor in real time. This allows for the generation of high-resolution maps of the whole field.

### 1.2 Related Work

Multispectral cameras have been used for some time to calculate vegetation indices. They have regularly been mounted on drones to survey agricultural land (Yu et al. (2025)). Alternatively, they can also be used on satellites to cover large areas. Projects like Sentinel-2 provide multi-spectral satellite images of Earth for free (CDSE (2015)). The resulting resolution is of course much lower than the drone-mounted variation. Additionally, cloud cover can block the view. Using vegetation sensors on ground vehicles to optimize **NUE** has similarly been discussed for some time. Ostermeier et al. (2006) already presented a strategy using the Red-Edge Inflection Point (**REIP**) index, based on ideas from Auernhammer et al. (1999). Remote and proximal sensing capture similar trends, but the accuracy tends to be better as the sensor is deployed closer to the target (Mezera et al. (2021)).

The Augmenta Field Analyzer (Augmenta (2019)) claims to have the ability to generate maps of **NDVI** and other indices from multispectral cameras while mounted to a tractor. It has already been used in literature to test variable rate application (**VRA**) setups. The system has a 5-spectrum camera system and multiple daylight sensors to generate a high-resolution **NDVI** map (Silvestri et al. (2024)). No detailed descriptions have been found on the working principle of the Augmenta system.

⋆ The project "EcoSchemeN" is supported by funds of the Federal Ministry of Food and Agriculture (BMEL) based on a decision of the Parliament of the Federal Republic of Germany via the Federal Office for Agriculture and Food (BLE) under the strategy for digitalisation in agriculture.

Multispectral cameras can still be expensive. However, approaches do exist that may only require cheaply available RGB cameras. Some vegetation indices can be calculated directly from RGB data. Fuentes-Peailillo et al. (2018) compared some of these to traditional indices like **NDVI**. While the new indices are not as good as the traditional ones, they still show useful results. Davidson et al. (2022) even shows an approach to estimate **NDVI** from RGB images using deep learning techniques.

## 2. METHODS & MATERIALS

### *2.1 Applied Methods*

To estimate the health and status of a plant, various vegetation indices have been introduced. The approach presented in this paper is limited to indices that can be calculated from the wavelengths measured by the used camera. With the device used in this work (see Section 2.2) Normalized Difference Vegetation Index (**NDVI**) and Normalized Difference Red Edge Index (**NDRE**) are available. For simplicity, all experiments are only shown for the **NDVI**. The **NDVI** was first introduced in its current form by Rouse et al. (1974). It compares the reflectance of plants in the red spectrum to their reflectance in the near-infrared (**NIR**) spectrum. The value is normalized by the sum of the reflectances, leading to (1).

$$\mathrm{NDVI} = \frac{r_{red} - r_{nir}}{r_{red} + r_{nir}} \tag{1}$$

A value near 0 indicates soil or dead plants, while a value near 1 indicates highly healthy plants.

For each image collected with the cameras, the corresponding position on the ground needs to be calculated. This is done using perspective projection, also referred to as backprojection. A simple pinhole camera model with a principal point offset was employed, as described in Hartley & Zisserman (2011) using a camera matrix $K$:

$$K = \begin{pmatrix} f & 0 & x_0 \\ 0 & f & y_0 \\ 0 & 0 & 1 \end{pmatrix} \tag{2}$$

where f is the focal length of the camera in pixels and $x_0, y_0$ are the x and y position of the image center in pixels. From this information, the local camera rays $t_{i,l}$ for all pixels (also referred to as the preimage) can be computed:

$$t_{i,l} = \begin{pmatrix} u_i \\ v_i \\ 1 \end{pmatrix} K^{-1} \tag{3}$$

where $u_i, v_i$ are the positions in pixels of each pixel.

The preimage then needs to be transformed into the global coordinate frame. This is done using standard rotation matrices and is therefore not further explained. The global preimage rays are referred to as $t_i$.

The preimage only represents the directional rays of each pixel from the camera origin. For each ray, the distance of the imaged object to that origin also needs to be recreated to form the scene. In this project the imaged object is always the ground. Therefore the intersection of the preimage and the ground plane defines the scene, giving a projected 3D position $p_i$ of each pixel in the global frame:

$$p_i = p_{cam} + \frac{p_{cam,z}}{t_{i,z}} r_i \tag{4}$$

where $p_{cam}$ is the position of the camera origin in the global frame and the subscript $,z$ indicates the z component of a vector. This approach assumes a flat ground within the area that is seen by the camera. More exactly, the ground is assumed flat relative to the attitude of the vehicle. Since the tractors attitude is parallel to the ground it stands on, this is a good estimation. The assumption gets worse for strongly curved fields and for long distances. To limit this problem, only points within 20m of the camera were used in later stages.

### *2.2 Used Material*

As a multispectral camera the Double 4k NDVI/NDRE Version (Sentera Inc, USA) was used (Sentera (2025)). It is usually carried by a drone and not originally designed for live image usage. However, by using the built-in HTML interfaces, it is possible to capture and download images from the device via Ethernet connection. It has a focal length of $f = 5.4$ mm and a horizontal field of view of 60°. The image size was typically set to 1024x768 pixels. This reduces transfer and computation times, while still providing sufficient resolution for the final map. The camera was mounted to an Aero fertilizer spreader (RAUCH Landmaschinenfabrik GmbH, Germany) in a fixed position. The spreader was used purely as a mounting point for the experiments mentioned in this paper. On the spreader, a dual antenna RTK-**GNSS** setup was also installed. This setup allows for precise measurement of the position and orientation of the implement. To also track the yaw rate, as well as the pitch and roll of the setup, an inertial measurement unit (**IMU**) (GEMAC Chemnitz GmbH, Germany) was installed. Having accurate estimation of the orientation is necessary for a good backprojection. A course over ground (**COG**) estimation, based on an extended kalman filter (**EKF**) fusing global navigation satellite system (**GNSS**) position and **IMU** data, was also tested in this project and found to be sufficient for this approach. The fertilizer spreader was hitched to a Fendt tractor. The positions of the sensors were precisely measured relative to the tractor rear axle (Fig. 1).

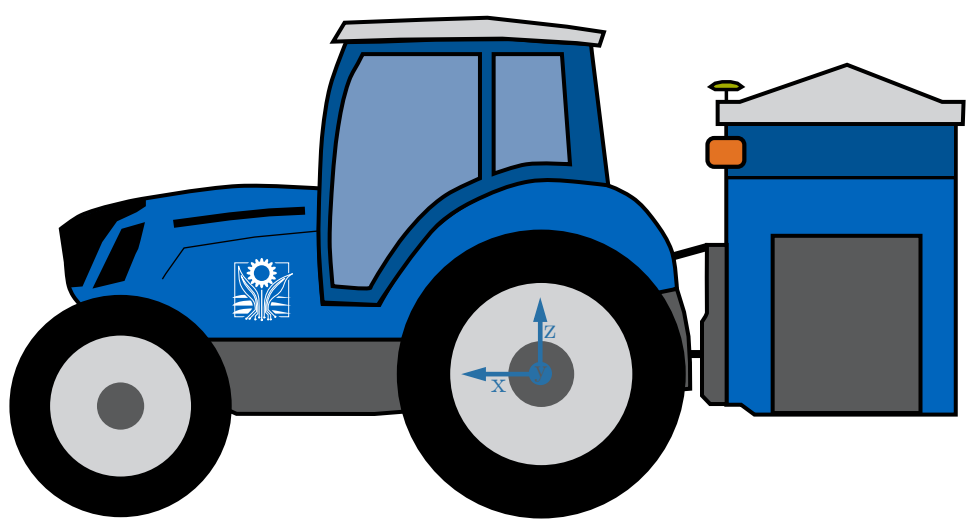

Fig. 1. Schematic overview of the physical setup. **GNSS** Antenna in green, camera in orange. The coordinate system is indicated in blue and centered in the rear axle of the tractor. The **PAR** sensor is placed with the camera.

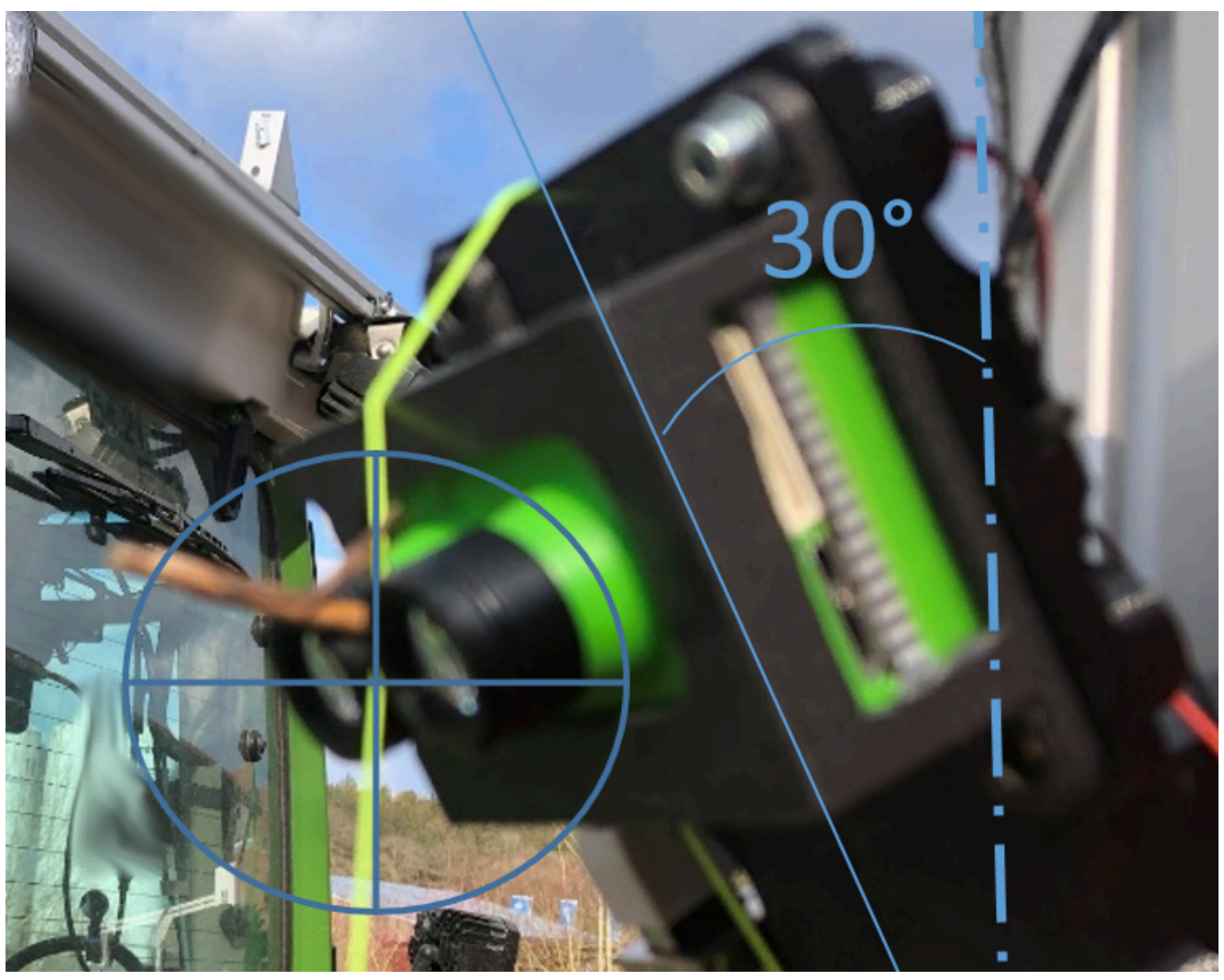

Fig. 2. The mounted camera.

The camera was mounted 30° from the upright position ($\varphi_y = 30°$) to focus it on the ground area. It was also rotated outwards, to look at the field instead of the tractor in front ($\varphi_z = 45°$). The positions of the camera and **GNSS** antennas in the local vehicle frame were: $p_{cam,v} = \begin{pmatrix}-1.015 & 1.415 & 1.36\end{pmatrix}^T$ m and $p_{gnss,v} = \begin{pmatrix}-1.085 & \pm 1.337 & 1.795\end{pmatrix}^T$ m. The rear axle and therefore the vehicle frame was measured to be 0.75m above the ground under normal conditions. A SQ-520 Photosynthetically Active Radiation (**PAR**) sensor (Apogee Instruments Inc, USA) was installed above the camera to track sunlight intensity, which might impact the readings from the camera.

A pair of commercial crop sensors "N-Sensor" (Yara International ASA, Norway) was also mounted to the spreader boom. They were pointed straight down, directly at the parcels measuring one data point per second each. From the measured data, the NDVI could also easily be calculated. These sensors were used only for comparison to the camera-based setup. They are known to be well calibrated and are in use for **VRA** worldwide.

A total of 40 parcels were used for testing of the system. Winter wheat was planted on all of them. Each individual parcel was 10m x 3m in size. Additional parcels were placed on the edge as a buffer to neighbouring plants. Ten different fertilization strategies were randomly placed per column. In the first application, which is used for the tests in this publication, only five different nitrogen amounts were given. See Fig. 3 for a drone image of the area. The parcel setup is explained in more detail in Pindl et al. (2026).

The algorithms were tested on two computers. The first one was a laptop with an Intel Core i5 CPU at 1.60 GHz, 16 GB RAM with no GPU support. The second one was a more powerful PC with an i9 CPU at 3.20 MHz, 32 GB RAM and a Nvidia GeForce RTX 4070 graphics card.

## 3. ALGORITHM

### *3.1 Image Acquisition and Processing*

To trigger an image, an HTML PUT request is sent via Ethernet. In multiple tests, there was no measurable delay between sending this request and the trigger time of the image. However, it may take over a second until the image is available on the camera. More images can be triggered in that time, but the first one cannot yet be received. The code is therefore multithreaded, so that the system can wait for each new image to be available while doing other jobs in the background. A list of trigger times for each image is kept, so this time can later be used for exact position calculations. Once the camera indicates that the new image is available, it is downloaded via an HTML GET request. The camera system also includes a built-in **IMU**. However, the time synchronization of this system seems to be too inaccurate. This data was therefore ignored.

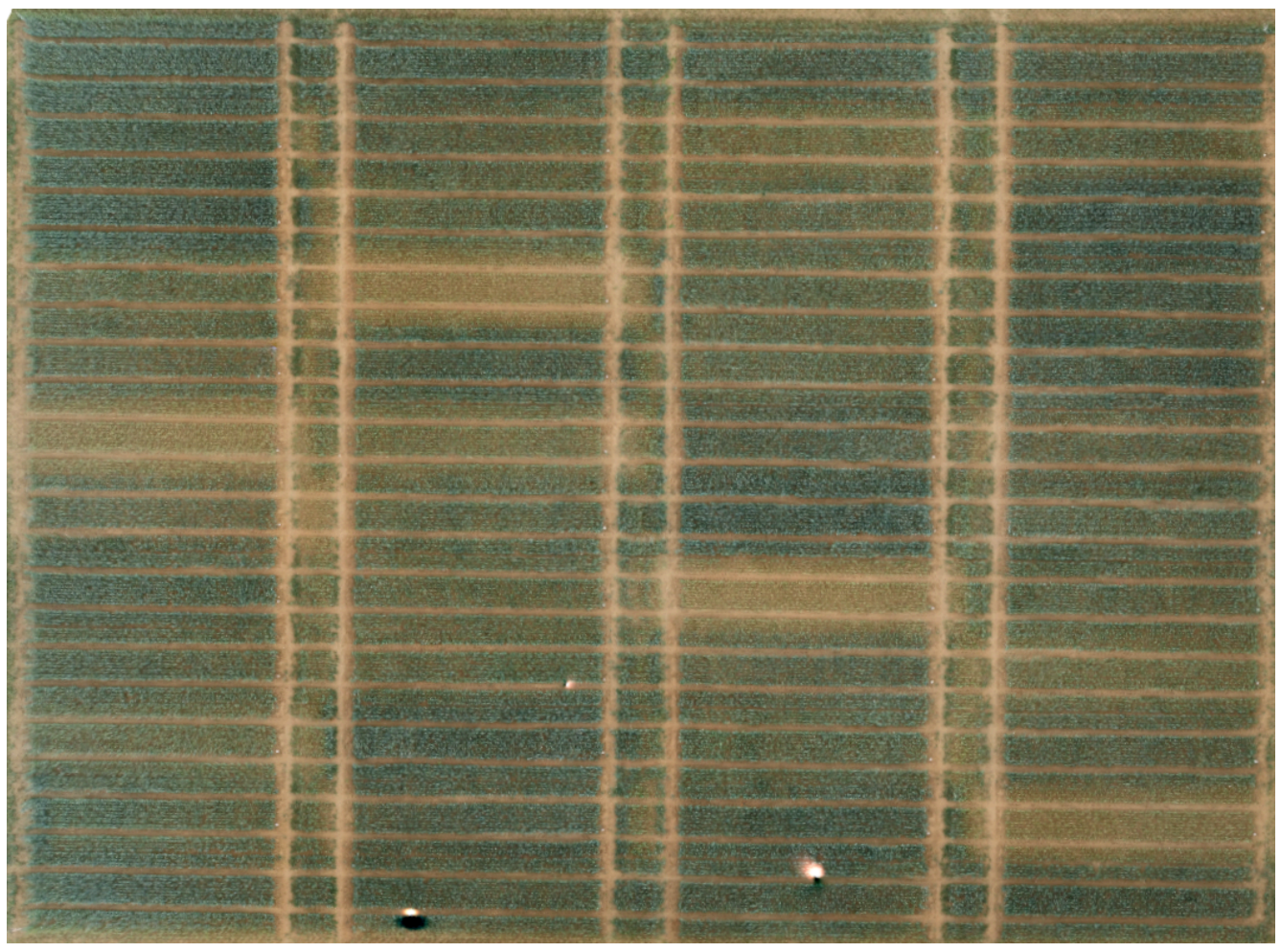

Fig. 3. A drone-based overview of the parcels, rotated from original orientation. The tractor drove up and down the vertical tramlines for data collection.

For each image the **NDVI** is calculated per pixel. Since the color filters used in the camera are not perfectly sharp, the reflectances need to be calculated according to the manual (Sentera (2025)):

$$\begin{aligned} r_{red} &= -0.966 d_{blue} + 1 d_{red} \\ r_{nir} &= 4.35 d_{blue} - 0.286 d_{red} \end{aligned} \tag{5}$$

where $d_{blue}, d_{red}$ are the data values for blue and red in the image file. While they do not actually represent blue and red color, the jpg format used by the camera assumes a typical RGB camera. The reflectance values are combined with (1) to calculate the **NDVI**.

(5) indicates that the different color channels do not have similarly scaled light responses. That means that a very bright white pixel may oversaturate the blue and red data at the same time ($d_{blue} = d_{red} = 1$), even though the underlying reflectances are also almost equal. Similar reflectances should mean an **NDVI** near 0, but the oversaturation leads to calculated reflectances of $r_{red} \approx 0, r_{nir} \approx 4$, meaning the system would register an **NDVI** around $-1$. Similarly, an **NDVI** close to 1 might be measured when only one filter is oversaturated, even though the actual value should be around 0. To filter out such instances, any pixel where the blue or red filter has a value close to its maximum is ignored in the analysis. A typical threshold was $d > \frac{245}{256}$.

Since the camera only uses 8-bit color resolution, darker areas will have very noisy **NDVI** readings. This is because the sum of both reflectances (denominator of (1)) is very small, meaning even a minor change will significantly change the ratio that constitutes the **NDVI**. Therefore pixels were also ignored if both the blue and red channel were very small. A typical threshold was $d < \frac{25}{256}$.

Since the **NDVI** is later averaged over larger areas (see Section 3.2), this will average over pixels showing ground as well as plants. Depending on the exact application, it may be useful to include ground pixels, since they typically have a lower **NDVI**, meaning areas with lower biomass tend to have a lower **NDVI**. For other applications however, it may be necessary to only analyse the difference in **NDVI** of the actual plants. To enable this use case, an optional filter is applied at this stage that filters out any pixels with an **NDVI** value smaller than some threshold, meaning ground pixels are ignored. Any pixel that was marked as ignored in any of the mentioned filters is not added to the map in the next step.

### *3.2 Map Projection and Merging*

Each **NDVI** image from the previous section is then projected onto the ground as explained in Section 2.1. The global position of the camera $p_{cam}$ is calculated from the local camera position $p_{cam,v}$ and the estimated position of the vehicle. For very high plants, the z component of the camera position may be adjusted to account for plant height. Especially in dense growth, the effective ground plane is shifted to the top of the plants. Tests showed however that for most cases the approach was sufficiently accurate even without this adjustment. The vehicle position was estimated from **GNSS** and **IMU** measurements using an **EKF**. The new estimated position was saved with every sensor measurement. To get an even better estimation for the exact trigger time of each image, the closest saved **EKF** state was used as a basis and a prediction step was done with the remaining time difference. Since typical sensor rates were around 10Hz even for the **GNSS** data, the prediction horizons were minimal and the estimated positions were very accurate. If the projected position of a pixel was more than 20m away from the camera, or closer than 1m, it was discarded. Far away points tend to get inaccurate, while for very close points the angle can be problematic, so that the plants are barely visible.

To optimize the computation time of the actual projection, multiple steps were taken. The local rays of the camera $t_i, l$ must only be computed once and are then just rotated into the global frame for each position. This is significantly cheaper than directly computing the global rays each time. The most expensive remaining calculation is finding the intersection of each ray with the ground. The corresponding function is fully vectorised and precompiled using the Python numba extension. When a supported GPU is available, most computations are shifted to CUDA, significantly speeding them up.

The projected points are then placed on a grid map with a predefined cell size, side length and center position. Points outside the map area are ignored. A cell size of 20cm was used. All points within one grid cell are averaged and the grid cell is filled with the average. The resulting map is merged with the existing map by filling cells that were previously empty. Cells that are filled in both the old and new map are combined with a weighted average:

$$g_{merged} = (1 - k)g_{old} + kg_{new} \tag{6}$$

where $k$ is a weight between 0 and 1, indicating how much of the new value should be used compared to the old. All code is set up to support multiple cameras merged on a single map. So far only one camera has been available however, meaning no testing with multiple sources has been done.

To analyse the data in the map, each parcel is checked separately. All pixels that are placed within the same parcel are averaged to get a mean **NDVI** for each parcel. Using the known nitrogen application amounts, the effects of different fertilization strategies can be compared. This is done similarly for the data from the commercial plant sensor to place each measurement into a parcel. Since the edge of the parcels is expected to be a transitional area, the actually analysed area is reduced from all sides by a buffer size. Usually a value of 0.8m was used.

## 4. DATA COLLECTION

All data shown in this work was collected on a sunny day at the end of April. This was about one month after the parcels were fertilized for the first time. The tractor drove up the rightmost tramline first (see Fig. 3), then down on the left tramline. Afterwards it turned around and followed the same path back. This allowed the camera to cover all parcels, even though it only looked towards the left.

## 5. RESULTS

Fig. 4 shows the final generated map for parcels. While the overall map was slightly bigger, the shown map was cut to better highlight the areas of interest. The shown map did not have a filter applied to ignore ground pixels.

The results of the map analysis are displayed in Fig. 5. The **NDVI** clearly rises from 0 to 40 kg/ha of nitrogen. At higher amounts the curve slowly flattens out. Each blue point is the mean value for one parcel orange points are the average per nitrogen step. Bars indicate the standard deviation of the data for each nitrogen step.

As expected, when a filter is activated to ignore ground pixels, the average **NDVI** is higher (Fig. 6). The general shape and distribution are similar to the unfiltered case.

Finally, a comparison between the commercial sensor and the new camera-based system is made in Fig. 7. It is difficult to directly compare the absolute **NDVI** values of different sensor systems. This is partially because the used wavelengths may not be exactly the same, but also due to different calibration. However the correlation between both sensors is clearly visible with a pearson factor of 0.841 when using the mean of each parcel as datapoints. Both sensors also show a similar standard deviation.

The two mentioned computers (see Section 2.2) showed significantly different performance. The laptop system projected 459 images in 80.0s. A single image therefore took on average 174ms. The GPU-supported system could project the same images in 6.14s, leading to an average time of 13.4ms. These times include reading the images from disk, projecting them on the ground, and merging the new map with the old.

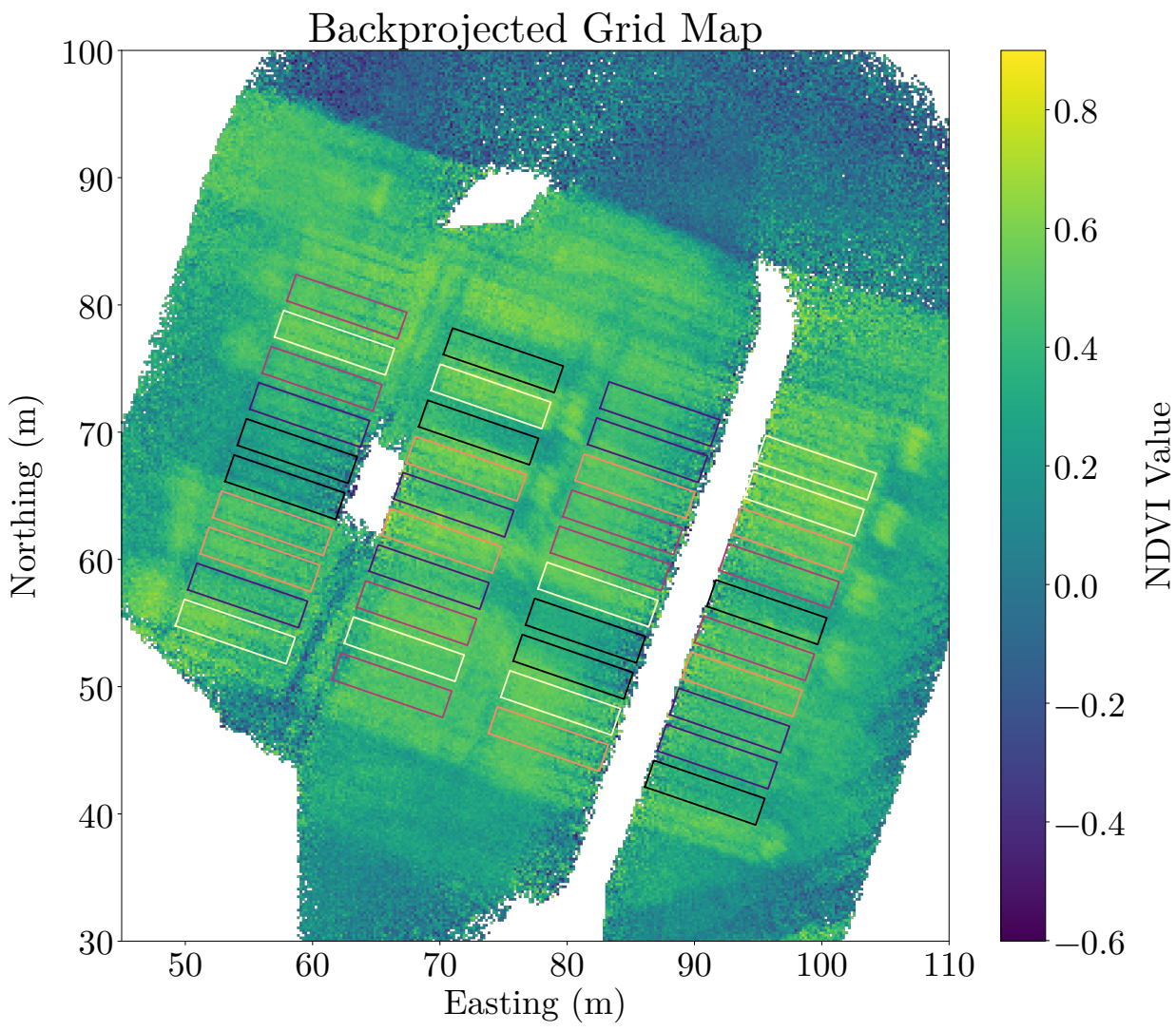

Fig. 4. The final **NDVI** map for a set of test parcels. The color of each parcel edge indicates the amount of nitrogen applied. Dark colors indicate low amounts, light colors indicate high amounts. A smaller buffer size was used for the parcels in this plot, to make them appear bigger purely for better visibility.

## 6. DISCUSSION AND CONCLUSION

The generated map (Fig. 4) visibly fits the laid-out parcels. Areas with higher fertilization are also brighter, indicating a higher measured **NDVI**. The measured nitrogen response (Fig. 5) is as expected. At lower amounts an increase has a high impact. At higher amounts other factors limit the growth more, leading to a plateau in the curve. The filter for ground pixels does not seem to have a major impact for the data in this instance (Fig. 6), but it may be useful for certain applications. It allows separation of the effects of visible plant coverage from the differences in **NDVI** of the plants themselves. The comparison to the commercial sensor (Fig. 7) is difficult since different sensors are not expected to measure the same absolute values. However, the shape of both curves is very similar, indicating that they capture the same trends. The standard deviation is also similar. This indicates that the relatively high variance may partially stem from variation in the actual growth between the different parcel repetitions and may not come from inaccuracies in the sensor. The projection times are fast enough, even on the laptop setup, that they could run for two cameras at 1Hz each, which is roughly the limit for the used camera model. Using another programming language instead of Python, this could likely still be improved. Overall the planned system was implemented successfully. Multispectral cameras necessary for this approach are still quite expensive. However, with increased use, prices may go down. Additionally, there is promising research to use RGB cameras instead, as discussed in the introduction. Over the next growing seasons, more data needs to be collected with this new sensor to test performance. Effects like the impact of sunlight intensity may then be tracked with the additional sensors that have been added to the system. Special care should be taken to further investigate the high variance that was found in the measured nitrogen response. The projection of the image can be further improved. Radial distortion was not taken into account for this work so far. The ground plane was assumed to be perfectly flat. If more information is available, for example from other sensors, a better estimation could be done.

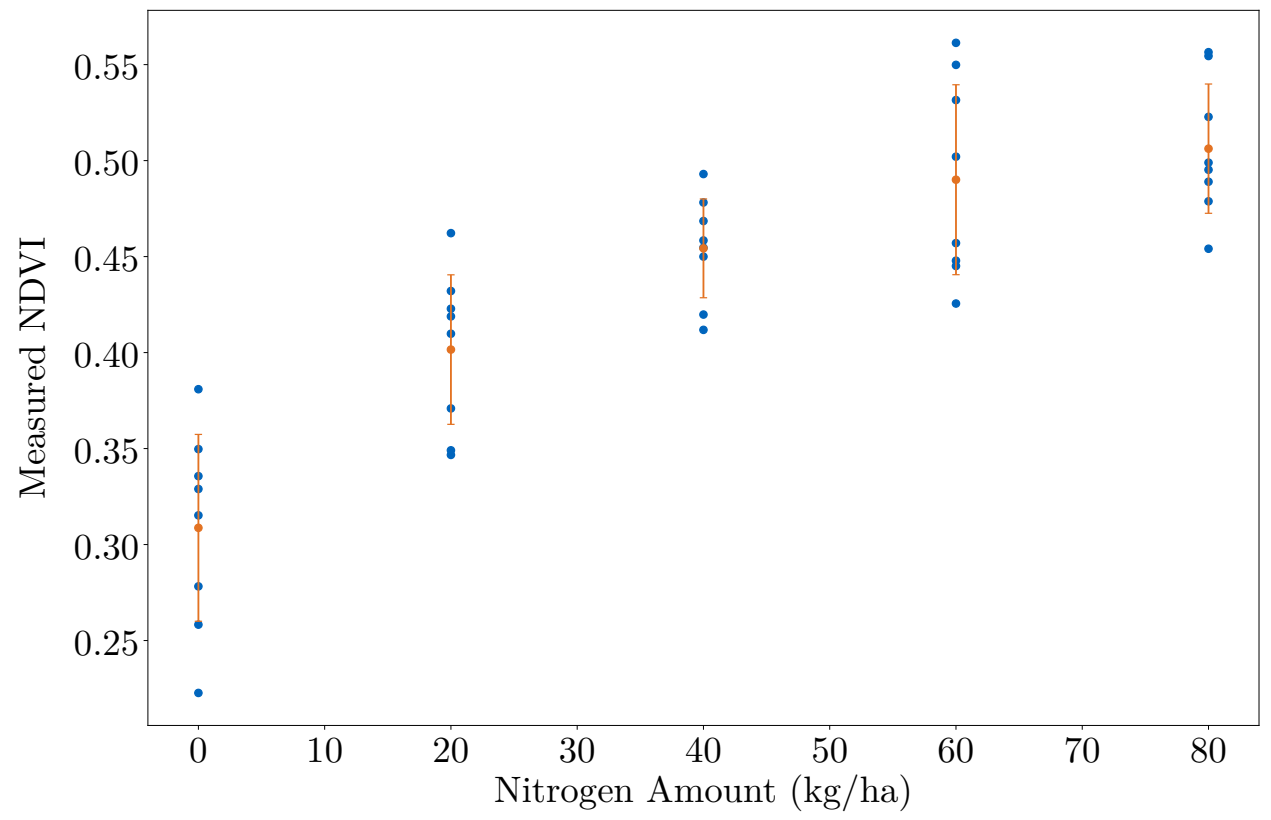

Fig. 5. The measured NDVI compared at different applied nitrogen amounts. Each blue point represents the mean for one parcel.

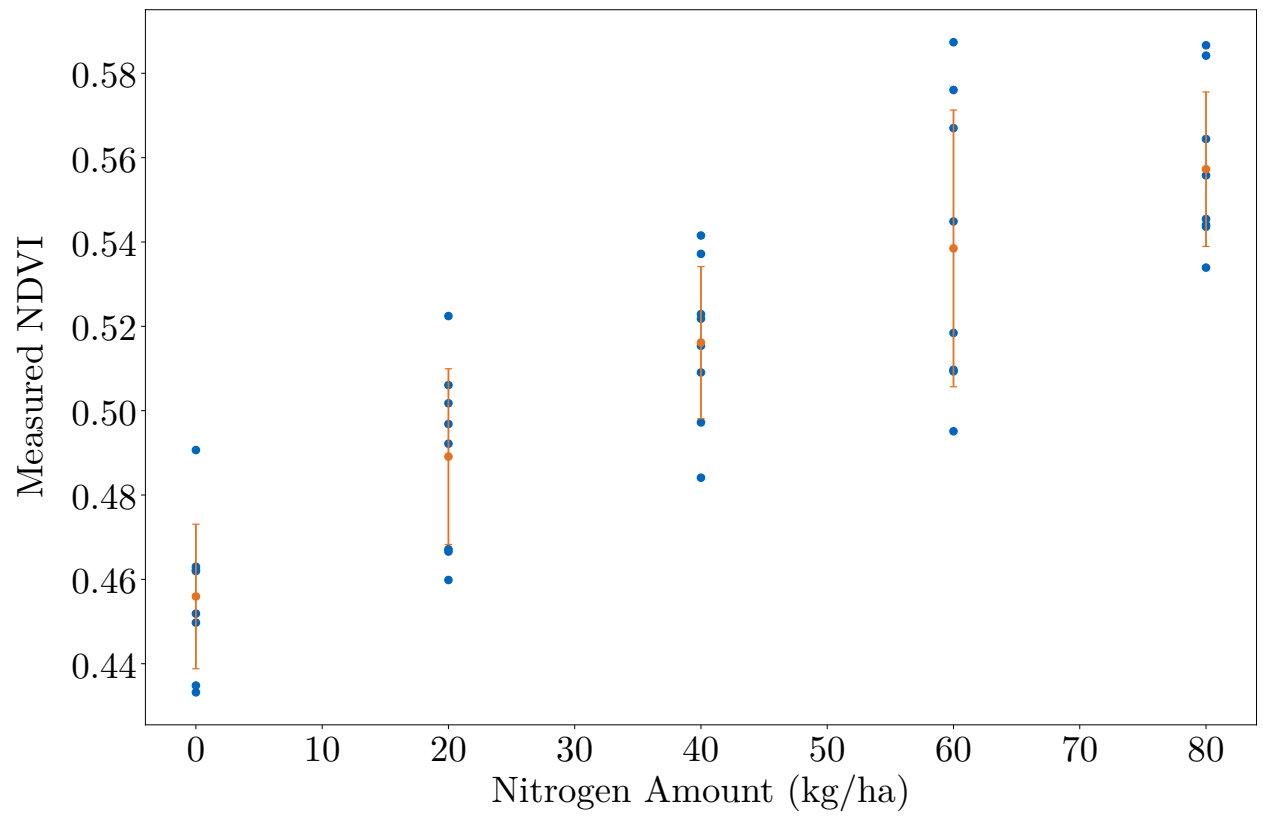

Fig. 6. The same comparison as in Fig. 5, but with ground pixels ignored. Only pixels with an **NDVI** value $> 0.3$ were used to generate the map in this case.

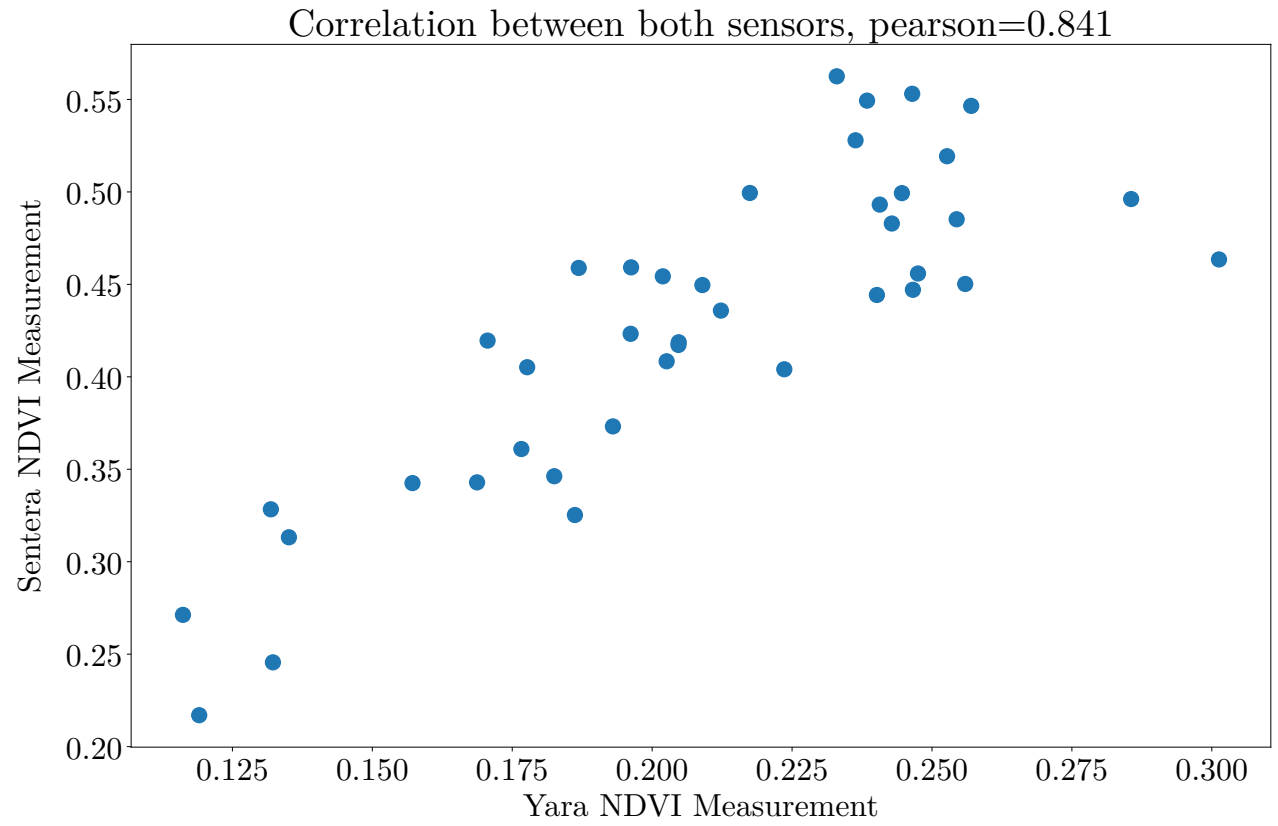

Fig. 7. Comparison between data from the commercial sensor (Yara) and the new camera-based approach (Sentera). Shows the mean for each of the 40 parcels.

## ACKNOWLEDGMENTS

We want to thank Yara for giving us access to additional software to get more detailed data from their sensors, which was used for comparison in this paper. We also would like to thank Rauch for the cooperation in this project and for making a spreader available to us, on which most sensors were mounted. The authors acknowledge the facilities, as well as the scientific and technical support, provided by the Field Crop Unit of the Plant Technology Center (PTC) at the Technical University of Munich (TUM).

## DECLARATION OF GENERATIVE AI AND AI-ASSISTED TECHNOLOGIES IN THE WRITING PROCESS

During the preparation of this work the authors used Microsoft Copilot in order to correct spelling and grammar mistakes. After using this tool/service, the authors reviewed and edited the content as needed and take full responsibility for the content of the publication.

## REFERENCES

Auernhammer, H., Demmel, M., Maidl, F. X., Schmidhalter, U., Schneider, T., & Wagner, P. (1999). An on-farm communication system for precision farming with nitrogen real-time application. *1999 ASAE/CSAE-SCGR Annual International Meeting.*

Augmenta. (2019). *Augmenta White Paper.* https://fanext.com/wp-content/uploads/2019/05/17-40-1-Augmenta.pdf

Barnes, E. M., Clarke, T. R., Richards, S. E., Colaizzi, P. D., Haberland, J., Kostrzewski, M., Lascano, R. J., & others. (2000). Coincident detection of crop water stress, nitrogen status and canopy density using ground-based multispectral data. *Proceedings of the 5th International Conference on Precision Agriculture.*

CDSE. (2015). *Sentinel-2 Documentation.* https://documentation.dataspace.copernicus.eu/Data/SentinelMissions/Sentinel2.html

Davidson, C., Jaganathan, V., Sivakumar, A. N., Czarnecki, J. M. P., & Chowdhary, G. (2022). NDVI/NDRE prediction from standard RGB aerial imagery using deep learning. *Computers and Electronics in Agriculture*, *203*, 107396. https://doi.org/10.1016/j.compag.2022.107396

EPA. (2019). *Memorandum of Understanding on Nutrient Science and Innovation and Watershed Approaches.* https://www.epa.gov/sites/default/files/2019-02/documents/mou_nsiw.pdf

Fuentes-Peailillo, F., Ortega-Farias, S., Rivera, M., Bardeen, M., & Moreno, M. (2018). Comparison of vegetation indices acquired from RGB and Multispectral sensors placed on UAV. *2018 IEEE International Conference on Automation/xxiii Congress of the Chilean Association of Automatic Control (ICA-ACCA)*, 1–6. https://doi.org/10.1109/ICA-ACCA.2018.8609861

Geupel, M., Heldstab, J., Schäppi, B., Reutimann, J., Bach, M., Häußermann, U., Knoll, L., Klement, L., & Breuer, L. (2021). A National Nitrogen Target for Germany. *Sustainability*, *13*(3), 1121. https://doi.org/10.3390/su13031121

Hartley, R., & Zisserman, A. (2011). *Multiple View Geometry in Computer Vision.* Cambridge University Press. https://doi.org/10.1017/CBO9780511811685

Mezera, J., Lukas, V., Horniaček, I., Smutný, V., & Elbl, J. (2021). Comparison of Proximal and Remote Sensing for the Diagnosis of Crop Status in Site-Specific Crop Management. *Sensors (Basel, Switzerland)*, *22*(1). https://doi.org/10.3390/s22010019

OpenDroneMap Authors. (2020). *ODM.* https://github.com/OpenDroneMap/ODM

Ostermeier, R., Rogge, H.-I., & Auernhammer, H. (2006). Multisensor Data Fusion Implementation for a Sensor based Fertilizer Application System. In M. Rothmund, M. Ehrl, & H. Auernhammer (Eds.), *Automation Technology for Off-Road Equipment* (pp. 215–225). Eigenverlag.

Pindl, L., Apfelböck, J., Maier, M., & Oksanen, T. (2026). *A Multimodal RGB and N-Sensor Dataset for Nitrogen Assessment in Winter Wheat* [Forschungsdaten].

Rouse, J. W., Haas, R. H., Schell, J. A., & Deering, D. W. (1974). Monitoring vegetation systems in the Great Plains with ERTS. *Proceedings, 3rd Earth Resources Technology Satellite (ERTS) Symposium*, *1*, 48–62.

Sentera (Ed.). (2025). *Sentera Double 4k Support.* https://support.sentera.com/portal/en/kb/articles/double-4k

Shen, W., He, J., Li, S., Zhuang, Y., Wang, H., Liu, H., Zhang, L., & Kappler, A. (2024). Opportunity and shift of nitrogen use in China. *Geography and Sustainability*, *5*(1), 33–40. https://doi.org/10.1016/j.geosus.2023.09.003

Silvestri, N., Ercolini, L., Grossi, N., & Ruggeri, M. (2024). Integrating NDVI and agronomic data to optimize the variable-rate nitrogen fertilization. *Precision Agriculture*, *25*(5), 2554–2572. https://doi.org/10.1007/s11119-024-10185-2

Yu, K., Belwalkar, A., Wang, W., Hu, Y., Hunegnaw, A., Nurunnabi, A., Ruf, T., Li, F., Jia, L., Kooistra, L., Miao, Y., & Teferle, F. N. (2025). UAV hyperspectral remote sensing for crop nitrogen monitoring: progress, challenges, and perspectives. *Smart Agricultural Technology*, *12*, 101507. https://doi.org/10.1016/j.atech.2025.101507